# Neural integrator - a sandpile model.

Maxim Nikitchenko and Alexei Koulakov

*Cold Spring Harbor Laboratory, Cold Spring Harbor, NY 11724*

**Abstract**

**We investigated a model for the neural integrator based on hysteretic units connected by positive feedback. Hysteresis is assumed to emerge from the intrinsic properties of the cells. We consider the recurrent networks containing either bistable or multistable neurons. We apply our analysis to the oculomotor velocity-to-position neural integrator that calculates the eye positions from the inputs that carry information about eye angular velocity. Using the analysis of the system in the parameter space we show the following. The direction of hysteresis in the neuronal response may be reversed for the system with recurrent connections compared to the case of unconnected neurons. Thus, for the NMDA receptor based bistability the firing rates after ON saccades may be higher than after OFF saccades for the same eye position. We suggest that this is an emergent property due to the presence of global recurrent feedback. The reversal of hysteresis occurs only when the size of hysteresis differs from neuron to neuron. We also relate the macroscopic leak time-constant of the integrator to the rate of microscopic spontaneous noise-driven transitions in the hysteretic units. Finally, we argue that the presence of neurons with small hysteresis may remove the threshold for integration.**

## Introduction

Persistent firing is a likely neural correlate of short-term memory[1, 2]. In some cases the variables stored in memory are continuous in nature[3]. Examples of such quantities include continuous sensory inputs[3-5], tension of a muscle, or variables representing accumulated sensory evidence[6, 7]. The continuously varying parameters are encoded in persistent neuronal firing, which has a graded set of values. The components of the nervous system that encode the graded values of parameters are called parametric memory systems [3-5].

Perhaps the best-studied system of this type is the oculomotor neural integrator[8-11]. Graded persistent activity in this system represents continuously varying eye position, which depends on the prior inputs carrying information about eye angular velocity. Since the transformation from velocity to position involves temporal integration, this system is also sometimes called velocity-to-position neural integrator (VPNI). The graded persistent activity in VPNI is likely to be maintained by positive feedback[10-13]. The presence of positive feedback poses a basic problem of robustness[10, 14]. This is because mistuning of the feedback leads to instabilities, which are hard to avoid in realistic systems. Previous researchers proposed that robustness could stem from hysteresis into neuronal responses[15, 16]. In this approach the robustness of networks based on hysteretic neurons is in many respects similar to the stability of digital electronic systems to mistuning of the parameters and noise.

A recent study in the goldfish oculomotor integrator[17] directly tested the history dependence in the responses of VPNI neurons. This study makes the following observations. First, the firing rate of a single neuron as a function of eye position exhibits hysteresis (Figure 1). Second, the firing rates during fixations are typically higher after the ON saccades than after the OFF saccades (Figure 1). This implies that the hysteresis in this system has an inverted direction compared to a typical positive feedback system, such as that due to nonlinear conductance of the NMDA receptor current (cf. e.g. Ref. [16, 18]). Third, the firing rate of one cell versus the other also displays history dependence. Fourth, the hysteresis width varies from cell to cell, with some cells showing no statistically substantial history dependence (Figure 1A).

Our present study addresses these experimental observations. We developed a simple model for VPNI that can be solved exactly without the use of a computer. We considered two related versions of this model, involving bistable and multistable neurons. The bistability is attributed to the bistable compartments within a single neuron[16, 19], while the multistability is formed by many bistable dendritic compartments. Although the specific mechanism is proposed, the properties of neurons in this model could be understood phenomenologically, and, perhaps, could be generated by many other possible intracellular or network mechanisms. The essential feature of our model, which allows it to be exactly solvable, is that the connectivity between neurons is all-to-all. This implies that all neurons receive the same value of input. We also present the results of a more biologically plausible computational model, which are consistent with the simpler solutions.

The main results of our study are as follows. First, we show that if the neurons in the absence of recurrent connections have hysteresis of regular sign[16, 19], adding global recurrent feedback produces the reversed hysteresis that is consistent with the higher firing rates seen after the ON saccades as described above. Thus, the direction of hysteresis observed experimentally could be attributed to the global recurrent connections between cells. Second, the phenomenon of the reversal of the sign of hysteresis occurs only if different neurons have different widths of hysteresis. Thus the experimental observation number two, that the firing rates are higher after the ON saccades, may follow from observation number four, that the hysteresis width varies from cell to cell. Finally, we studied the temporal properties of VPNI using a kinetic equation formulated in the parameter space of the system. We show that the rate of integration is controlled by the synaptic time constant $\tau_s$, which, in the case of the NMDA receptor is about 0.1 sec. On the other hand, the integrator leak time constant $\tau_{leak}$ is determined by the rate of spontaneous transitions in the bistable neurons denoted here $\tau_h$. The expression for the integrator leak is of the form

$$\tau_{leak} = \tau_h / \varepsilon. \qquad (1)$$

The parameter $\varepsilon \ll 1$ defines the precision with which the integrator is tuned. For the VPNI without hysteresis, the leak is given by the same expression with $\tau_h$ replaced by $\tau_s \sim 0.1\sec$ [10, 13]. Because the time-constant of spontaneous transitions $\tau_h$ is usually much larger than the synaptic time-constant $\tau_s \sim 0.1\sec$ [20, 21], the use of hysteretic neurons allows stabilization of the integrator at a much larger value of the precision of tuning $\varepsilon$, which provides another argument for the robustness of the hysteretic system.

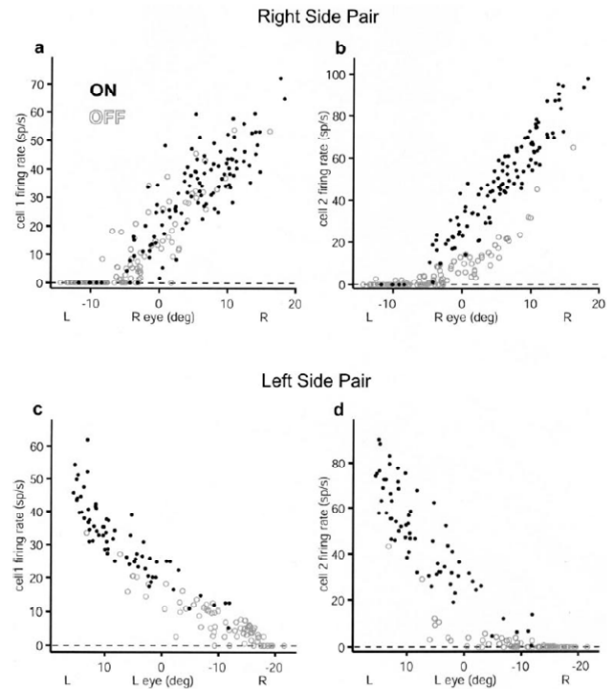

**Figure 1.** Responses of neurons in the goldfish medulla, area I exhibit hysteresis[17]. The firing rates after ON fixations are above the OFF fixations.

## Computational Model

The network model we used in the computational part of the study is similar to the previously described NMDA-based models[16, 19]. The network included 40 two-compartmental neurons. Each neuron contained the somatic and dendritic compartments. The somatic compartment included sodium and potassium currents making it capable of generating action potentials. The dendritic compartments received feedforward NMDA current, feedback NMDA current, and an offset current needed to distribute the thresholds for activation. The NMDA-based bistability was produced by the feedforward NMDA currents into the dendritic compartments. This current was due to feedforward inputs from 100 neurons discharging at 30Hz. The NMDA conductance for feedforward inputs was equal to 0.7, 1, 1.1, and 1.2 $\mu s/cm^2$ for neurons from groups 1 through 4 in Figure 11A. Different values of hysteresis between these neurons resulted from differing NMDA conductance. These four groups of neurons also received unspecified feedforward input currents of 1.98, 1.20, 0.93, and 0.68 $\mu A/cm^2$ to equate their mean thresholds for activation (parameter $\theta$) as shown in Figure 11A. To produce a difference in the mean thresholds, another offset current was added to the quadruples of neurons. Each quadruple was therefore separated by 0.05 $\mu A/cm^2$ (Figure 11A) in $\theta$-space from its nearest neighbor. The offset feedforward current needed to satisfy these assumptions could be of AMPA origin. However, no specific implementation for the synaptic current was introduced to simplify the numeric algorithm. The feedback connections between neurons contained NMDA conductance only of $4\,ns/cm^2$ (no AMPA current) for simplicity. For details of implementing these conductances see Refs. [16, 19]. The somatic, dendritic capacitances and leak current (both somatic and dendritic) were taken to be $1\,\mu F/cm^2$, $0.5\,\mu F/cm^2$, and 0.1 $mS/cm^2$ respectively[16, 19].

## Results

Our model for the neural integrator is based on recurrent positive feedback[10, 12, 13, 15, 16]. First we present the results obtained for a simplified model, which can be solved exactly without the use of a computer. To make the exact solution possible some assumptions have to be made about the recurrent network connectivity. The main assumption is that the neurons are connected in the all-to-all fashion with equal weights (Figure 2). In this case all neurons receive the same input current, which greatly simplifies the analysis. This assumption about network connectivity is in contrast to the one made by Goldman et. al., [Ref. [15]], who considered feedback connectivity targeting specific dendritic compartments.

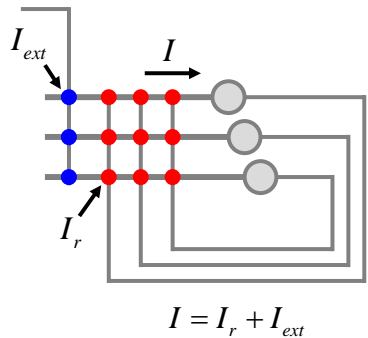

$$I = I_r + I_{ext}$$

**Figure 2.** The recurrent feedback model, which is thought to underlie the neural integrator. Recurrent synapses, external synapses, and somata are shown by red, blue, and gray circles respectively. In the fully-connected network considered here, all the neurons receive the same input (*I*). The input is a sum of external and recurrent currents ($I_{ext}$ and $I_r$ respectively).

### 1. Bistable neurons

## 1.1 The case of no recurrent feedback.

In this subsection we consider the properties of integrator neurons without recurrent connections. The recurrent connections are included in the subsection 1.2.

Figure 3A shows the response of the neuron that is used in this simplified model. The red and blue curves show the firing rate dependences for increasing and decreasing inputs respectively. If the input is in the range marked by the green arrow (region II in Figure 3A) the firing rate of the neuron may have two values depending on the prior history. The neuron is therefore bistable for the range of inputs indicated by the green arrow. If the neuron is in the state with the higher firing rate in the bistable regime, it is considered to be active or ON. The state with the lower firing rate is defined as deactivated or OFF state.

The response of a neuron as a function of input current exhibits three regimes shown in Figure 3A. For the values of input current below the bistable regime (region I in Figure 3A) the neuron is considered deactivated (OFF) unconditionally. For the inputs above the bistable range (region III), the neuron is considered always active (ON). Although in this simplified model the OFF state coincides with a firing rate equal to zero (Figure 3A), this assumption is made to make analysis of the model easier. The non-zero firing rates are possible in the OFF state if they do not activate NMDA postsynaptic currents substantially, as described below. The feature that is important for the subsequent analysis is that neurons display history dependence in their firing rate, which implies that the neuronal response exhibits hysteresis.

We consider an ensemble of units which differ in two respects: hysteresis width and position. The former parameter is described by the half-width of hysteresis $\Delta = (\theta^\uparrow - \theta^\downarrow)/2$, where $\theta^\uparrow$ and $\theta^\downarrow$ are thresholds for activation and deactivation respectively. The position of hysteresis is described by the average of two thresholds: $\theta = (\theta^\uparrow + \theta^\downarrow)/2$, as illustrated in Figure 3A. The ensemble of many of such neurons is distributed in the 2D parameter space $(\Delta, \theta)$ as shown in Figure 3B. The density of units in the parameter plane is

$$\rho(\Delta, \theta) = C \exp(-\Delta/\overline{\Delta}) \qquad (2)$$

Here, $C$ and $\overline{\Delta}$ are the normalization constant and the average half-width of hysteresis. The number of neurons in the square of parameter space with dimensions $d\Delta$ and $d\theta$ along the $\Delta$- and the $\theta$-axes respectively is given by $\rho(\Delta,\theta)d\Delta d\theta$ for a sufficiently small square. Thus, although $\rho$ depends only on one coordinate, it is a 2D density of neurons. This the 1D density along the $\theta$-axis is constant $\rho(\theta) = C\overline{\Delta} = \text{const}$. Although we adopted distribution (2) for concreteness, the analysis described below could be performed for an arbitrary distribution.

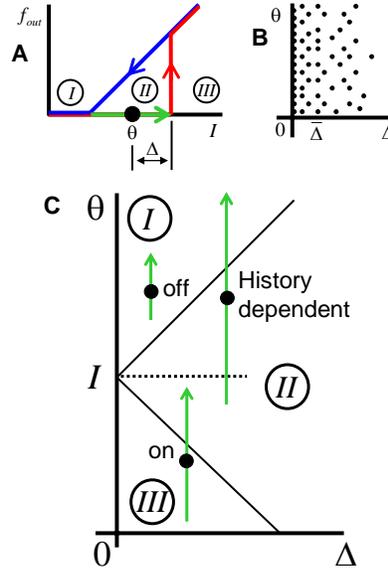

**Figure 3.** Ensemble of bistable neurons with differential parameters. (A) Example of the input to firing rate relationship for a bistable neuron. The activation (red) and deactivation curves (blue) do not coincide in the bistable range (green). If the input current (*I*) to the neuron is in the bistable range the firing rate can have two values depending on the previous history. The bistable range is shown by the green arrow, with the threshold for activation denoted by the arrowhead. The threshold for deactivation coincides with the arrow's tail. Such neurons can be described by two parameters: $\theta$ and $\Delta$, representing the mean position of the threshold (black dot) and the half-width of hysteresis respectively. The unconditionally OFF, bistable, and unconditionally ON ranges of input are labeled by I, II, and III respectively. (B) We considered an ensemble of many bistable neurons. Each neuron is represented by a point in the parameter space $(\Delta, \theta)$. The density of neurons is uniform along the vertical $\theta$-axis, while the density decays for lager values of the hysteresis half-width. (C) For a given value of the synaptic current on the inputs of all neurons (*I,* dotted line) the parameter space is divided into three regions. In the top region the neurons are unconditionally OFF. This region corresponds to the range of inputs labeled by I in (A). Thus for one of such neurons, whose bistable range is shown by the green arrow, the input level (dotted line) is below both thresholds for activation and deactivation, which implies that the neuron is unconditionally OFF. In the bottom region, labeled by III, all the neurons are unconditionally ON, since, as shown for another example neuron, the input is above the threshold for activation. In the middle region (labeled by II) the input current is in the bistable range for all neurons, as shown for one of the neurons. The firing state of these neurons is therefore history dependent, which implies that they can be either ON or OFF.

We now recall that in case of all-to-all connectivity all neurons receive the same value of synaptic input current (Figure 2). It is instructive therefore to consider properties of the neuronal ensemble when the same input is supplied to the neurons with different parameters. In particular it is of interest to determine what neurons are ON or OFF for a given value of input current. Clearly, an unambiguous answer to this question cannot be given. This is because for the given value of input current ($I$) there are neurons, which are in the bistable regime, i.e. their state depends on their history.

For a given value of input current all neurons can be divided into three groups: neurons, which are unconditionally OFF, ON, and the history-dependent units. In Figure 3C the areas occupied by these groups are marked by I, III, and II respectively. For neurons in these areas the input currents are in the ranges I, III, and II indicated in Figure 3A. For the neurons that are unconditionally ON (group III) the value of input current is above their threshold for activation, as follows from Figure 3A: $I \geq \theta^\uparrow = \theta + \Delta$. Therefore, such units are located in the region of the parameter space defined by the following condition:

$$\theta \leq I - \Delta \qquad (3)$$

The units which are unconditionally OFF (group I) receive input current, which is below their threshold for activation: $I \leq \theta^\downarrow = \theta - \Delta$. These units are therefore defined by another condition:

$$\theta \geq I + \Delta \qquad (4)$$

Finally, the neurons which are neither unconditionally ON nor OFF, have a state, which depends on history (group III). Their positions are defined by an alternative to (3) and (4):

$$I - \Delta \leq \theta \leq I + \Delta \qquad (5)$$

Note that the positions of areas I-III depend on the value of input current and therefore may change with time. By manipulating external current, one can form various patterns of activation and deactivation, some of which are shown in Figure 4. These patterns do not depend on the distribution of units in the parameter space, i.e. function $\rho$ in (2). The latter function only defined the density of neurons independently on whether they are ON or OFF. It is therefore *not* history dependent. Another distribution is needed to describe history dependence in the activation of hysteretic neurons.

We next define the *activation function* $h(\Delta, \theta)$. This function specifies if a neuron at a point with coordinates in the parameter space $(\Delta, \theta)$ is ON or OFF. It is equal to one in the areas occupied by active units and zero in other areas. The *total* number of active units in the ensemble is determined by the sum of the product of the densities over the parameter space:

$$n(t) = \int h(\Delta, \theta, t) \rho(\Delta, \theta) d\Delta d\theta \qquad (6)$$

In this equation the activation function $h$ acts as a marker, which allows inclusion of only the areas occupied by the active neurons in the sum. We emphasize that $n$ may depend on the history of prior inputs, since it includes the history-dependent activation function $h(\Delta, \theta, t)$. This expression will allow us to calculate the recurrent current in the next subsection.

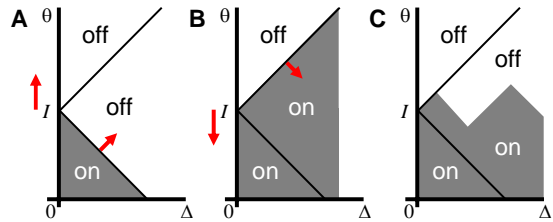

**Figure 4.** Possible configurations of active units (activation function $h(\Delta, \theta)$) on the absence of global recurrent feedback connections. The active areas are shaded. (A) If initially all neurons are OFF and the external current was increased from zero, only the units, which are unconditionally ON (group I) are active. (B) If initially all units are ON and the external current sweeps the parameters space downward, there is a deactivation wave propagating with the boundary between regions II and III (red slanted arrow). The active region extends indefinitely in the direction of a large $\Delta$ and is truncated in this and following figures. (C) A more complex pattern of activation in the history dependent region (III) can be produced by a complex pattern of inputs. For the profile shown, current was going up, down, up, and down.

**1.2 Recurrent feedback and the stability condition.**

In the previous subsection we studied the properties of a simplified model of hysteretic neurons with only feedforward connections present. Here we include the recurrent connections into the model. To this end we assume that the recurrent current $I_r$ is proportional to the total number of active neurons $n$ given by (6). We therefore neglect by the variations in the recurrent current due to changes in the neuronal firing rates assuming synaptic saturation[16, 22]. This approximation is valid, if a receptor with a large time constant, such as an NMDA receptor is responsible for neurotransmission in the recurrent synapses[16, 22]. The long time constant of NMDA receptors leads to saturation of synaptic currents even at small firing rates (10-20Hz), implying little dependence of the recurrent current on the firing rates. The saturation at low firing rates may also be of presynaptic nature, arising from synaptic depression, for example.

We will now address the dynamics of our model in the case of recurrent connections present. We relate the recurrent current to the number of active neurons given by (6). We will assume here that each active neuron contributes $I_0$ to the recurrent current. Therefore the total recurrent current into each neuron can be found as a product of the number of active neurons and parameter $I_0$

$$I_r(t) = I_0 n(t - \tau_s) \qquad (7)$$

Note that the recurrent current is related to the number of active neurons $n$ with a synaptic delay $\tau_s$. If the NMDA receptor is a primary neurotransmitter in the recurrent synapses, one should expect synaptic delay to be $\tau_s \simeq 100$ msec.

To complete the description of the simplified model we introduce the total value of the input current for each neuron (Figure 2)

$$I(t) = I_r(t) + I_{ext}(t) \qquad (8)$$

Here $I_{ext}$ is the external "command" input. The new value of input current $I(t)$ determines the neuronal activation function for the new time-step $h(\vec{p}, t)$ through the set of constraints (3)-(5). The activation function through equation (7) leads to a new value of recurrent current at the next time-step $t + \tau$. Thus the system of equations (3)-(8) allows accounting for iterative dynamics of the system of hysteretic neurons with different parameters connected by recurrent synapses. This dynamic is illustrated below on a series of examples.

We will first discuss the dynamics of this model in response to a tonic external input [$I_{ext}(t) = const$]. Let us first assume that the current is positive ($I_{ext} > 0$). It is expected then that under certain conditions, which become evident below, the input is integrated temporarily, implying that total current in the system increases with time. We then expect a wave of activation, similar to shown in Figure 4A to propagate upward in the parameter space. We now discuss the equations governing the propagation of this wave and the conditions of its existence.

Given the value of total current $I$, the number of active units is *approximately* given by the product of the area occupied by neurons under the dotted line in Figure 5A, $I\overline{\Delta}$, and the concentration of neurons in the parameter space $C$ [see (2)], i.e. $n = CI\overline{\Delta}$. A small correction has to be made to subtract the neurons, represented by black empty circles in Figure 5A. The corrected expression for the number of neurons in the ON state is

$$n(t) = C\left[I(t) - \overline{\Delta}\right]\overline{\Delta} \qquad (9)$$

The new value of current at the next time-step is, according to (7) and (8)

$$I(t + \tau_s) = \alpha I(t) + I_{ext} - \alpha \overline{\Delta} \qquad (10)$$

where we introduced the unitless parameter $\alpha = I_0 C \overline{\Delta}$. The "perfect integrator" condition corresponds to the value of parameter $\alpha = 1$. In this case the current is accumulated according to (10) without a loss:

$$I(t) = \left(I_{ext} - \overline{\Delta}\right) t / \tau_s + I(t = 0). \qquad (11)$$

The quantity being accumulated is $I_{ext} - \overline{\Delta}$. The system is therefore capable to act as a temporal integrator.

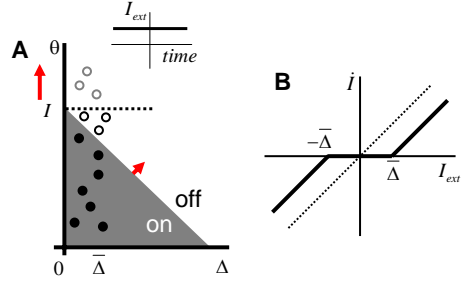

**Figure 5**. Response to tonic input.
(**A**) For non-zero external input that is constant in time (inset) the current in the system ($I$) may increase as a function of time (see **B** for the condition of this). In this case the number of active neurons (full circles) is given by (9). The inactive units are shown by empty circles.
(**B**) For stationary external input the system's response increases with time if the value of input exceeds a threshold equal to $\overline{\Delta}$. The rate of increase ($\dot{I} \equiv dI/dt$) is proportional to the external current above threshold.

It is however clear that sustained integration is possible only if external current exceeds the average value of hysteresis, i.e. $I_{ext} > \overline{\Delta}$ for $\alpha = 1$. For $I_{ext} < \overline{\Delta}$ the current would have to decrease according to (10), which is not valid in this case. Indeed, in deriving (11) we assumed that there is a wave of activation propagating up in the parameter space (Figure 5A), not down. Therefore sustained increase in the input synaptic current $I$ is not possible if $I_{ext} < \overline{\Delta}$. Consequently, the value of external input current of $\overline{\Delta}$ represents a threshold for integration. Similarly, only the negative inputs below $-\overline{\Delta}$ can be integrated in a sustained manner, which results in a negative threshold for integration. The rate of change in the synaptic current $\dot{I} \equiv dI/dt$ as a function of external input is summarized in Figure 5B. This figure shows that if the external input is between $-\overline{\Delta}$ and $\overline{\Delta}$, it is not integrated in a sustained manner. This statement is valid for an arbitrary distribution of active units in the parameter space $\rho$. That is, if, instead of an exponential distribution given by equation (2), any other distribution of hysteresis widths is found, the threshold for sustained integration is equal to the average value of hysteresis $\overline{\Delta}$. The issue of threshold for integration is discussed below is sections 4 and 5.

We now discuss the case of zero external input, which in the case of oculomotor VPI corresponds to eye fixations. It turns out that our model displays more sophisticated behaviors in this stationary condition rather than in the non-stationary one. The possible distributions of active units are shown in Figure 6. We start from the simplest case, in which the activation function is shown in Figure 6A. In this case all the neurons that have the medial threshold $\theta$ below the present value of input current $I$ are active. The interface between the ON and OFF neurons is a straight line parallel to the $\Delta$-axis. It is not so difficult to see that this distribution function cannot be realized using just the network architecture with all-to-all recurrent connections and the same synaptic input to all neurons. This is because in this architecture the boundaries separating ON and OFF neurons form a 45 degree angle with the parameter axes (Figures 4 and 5). Nevertheless, we will consider the activation function in Figure 6A, because it gives insights into more complex cases (Figure 6B and C). Assume that by manipulating each neuron individually the activation function in Figure 6A was set up. What is the condition needed for it to remain stable as a function of time?

To answer this question one has to repeat the simple calculations, which led us to equation (10). The number of active units as a function of total synaptic current is $n(t) = CI(t)\overline{\Delta}$ [see the text preceding equation (9)]. The value of the total synaptic current $I$ at the next time-step $t+\tau$ is equal to the total recurrent current since $I_{ext} = 0$:

$$I(t+\tau_s) = \alpha I(t) \qquad (12)$$

Comparing this equation with (10), we see that both the input current and the threshold equal to $\overline{\Delta}$ have disappeared from the equation. For the configuration in Figure 6A to remain stable one has to satisfy the condition of stationarity of the synaptic current, i.e. $I(t+\tau_s) = I(t)$. In view of equation (12) this is equivalent to setting $\alpha = 1$, i.e. having a perfectly tuned integrator. We conclude that the activation function in Figure 6A is stable for a perfectly tuned integrator ($\alpha = 1$). We will assume the perfect integration for the remainder of this subsection. Some solutions for a non-perfect integrator will be given in the next subsection.

What other stable activation functions are possible? One can generate stable activation functions from the previous example (Figure 6A). To achieve this, the following two conditions should be met. First, the number of active neurons should be the same as in Figure 6A. Thus, Figure 6B shows the activation function for which the number of newly recruited neurons (shaded red) is compensated by the deactivated ones (cyan), so that the total number of active neurons is the same as in Figure 6A. Second, the intersection of the boundary between ON and OFF neurons with the vertical axis $\theta$ should be the same. This is because this intersection determines the total synaptic current $I$, which is not perturbed according to the first requirement. The first requirement (the change in the number of ON neurons with respect to Figure 6A $\Delta n$ is zero) amounts to

$$\Delta n = \int_0^\infty b(\Delta)\rho(\Delta)d\Delta = 0 \qquad (13)$$

The boundary function $b(\Delta)$ is the shape the interface between the active and inactive neurons, which is set to be zero at $\Delta = 0$ (Figures 6D-C). This function describes the deviation of the shape of the interface from the case shown in Figure 6A. The boundary function is positive in the areas in Figure 6 having a shade of red and is negative in the cyan areas. Equation (13) implies that an *increase* in the number of active units in the red areas in Figures 6B and C, where $b(\Delta) > 0$, is compensated by the *decrease* in the cyan areas [$b(\Delta) < 0$], thus leading to *no* overall change in the total number of active units. If $\Delta n = 0$ the recurrent current is the same as in Figure 6A leading to the stable configuration. We refer to condition (13) as the *stability condition*. We will show below how this condition leads to the reversal of the sign of hysteresis in the firing rate as a function of eye-position dependence.

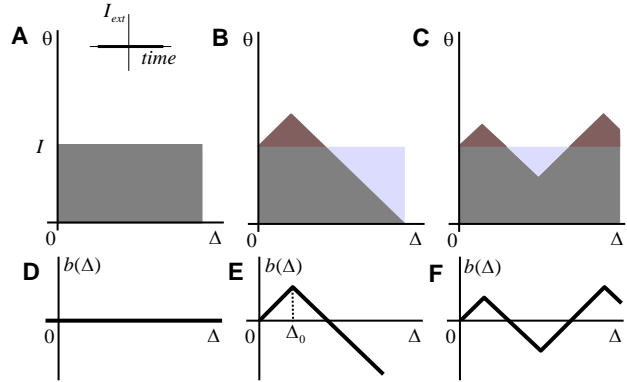

**Figure 6**. The case of zero external input (inset).
(**A-C**) Possible stationary configurations of the activation function. The areas shaded by red/cyan lead to an increase/decrease in the number of active neurons with respect to (A). Stable configurations are achieved when the red and cyan areas have the same number of neurons.
(**D-F**) The boundary functions $b(\Delta)$ corresponding to the activation profiles in (A-C).

The parameter $\Delta_0$ determines the position of the maximum of the function in Figure 6B (the wedge in the activation function). This parameter can be found from (13). In Appendix A we calculate $\Delta_0$ for the exponential distribution of cellular hystereses (2)

$$\Delta_0 = \overline{\Delta} \ln 2 \qquad (14)$$

This relationship is valid only for the exponential distribution of the hysteresis widths. For a different distribution different from exponential, a different coefficient of proportionality between $\Delta_0$ and $\overline{\Delta}$ is expected (Appendix A). Thus, previously we considered the ensemble of neurons with the same values of hysteresis width[16]. This ensemble is defined by the distribution that replaces (2):

$$\rho(\Delta) = C\overline{\Delta}\delta(\Delta - \overline{\Delta}) \qquad (15)$$

where $\delta$ is the Dirac delta-function. The stability condition (13) can also be used to calculate the parameter $\Delta_0$ for this distribution:

$$\Delta_0 = \overline{\Delta}/2 \qquad (16)$$

Thus, although the numerical coefficients for exponential and delta-function distributions differ between equations (14) and (16), the value of parameter $\Delta_0$ is determined by the average hysteresis width, $\overline{\Delta}$. Finally, the stability condition (13) allows finding stable parameters of more complex configurations, such as the one shown in Figure 6C. The stability condition allows formulating an equation for the positions of the vertex points of the activation function, such as the vertex at $\Delta = \Delta_0$ in Figure 6E.

We will now describe the transitional regime between integration ($I_{ext} > \Delta_0$, eye movement) and the stable configuration described in the previous paragraph ($I_{ext} = 0$, eye fixation). With the external current present, the activation function is described by the wave of activation propagating in the parameter space (Figures 5A reproduced in Figure 7A), which means that the eye position is increasing. Assume that the external current is suddenly removed (Figure 7A, inset). The activation function in Figure 7A is formed during eye movements and does not satisfy the stability condition. As such, it cannot exist during eye fixations. The activation function has to evolve to one of the stable configurations, such as that shown in Figure 6. In the simplest case the activation function evolves to the configuration in Figures 6B and C also shown in Figure 7B. As a result, the recurrent current $I$ drops after the removal of external current in the direction opposite to the eye movement. The amount of such a recurrent current drop $\Delta I$ is

$$\Delta I = 2\Delta_0 = 2\overline{\Delta}\ln 2 \qquad (17)$$

This drop is given here with a positive sign despite the fact that the current was decreasing after the disappearance of the external input.

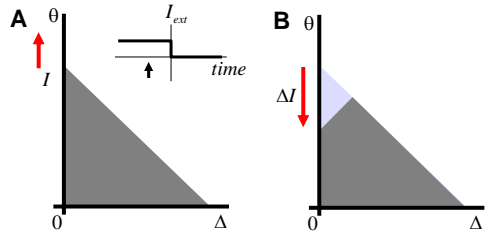

**Figure 7**. Response to input current that suddenly drops to zero (insets). The black arrows in the insets indicate the moment of time when this activation function is expected.

To calculate the firing rate as a function of eye position during eye fixations we will assume here that the eye position is proportional to the recurrent current. A more complicated relationship does not change our conclusions qualitatively. During eye fixations the external input is absent, i.e. $I_{ext} = 0$. The eye position is therefore equal to the total synaptic current $I$ [see (8)]. Ignoring the proportionality constant we assume that during fixations the eye position is equal to the input synaptic current

$$E = I . \qquad (18)$$

The problem of finding the firing rate as a function of eye position is therefore seemingly simple: to determine the response of a neuron as a function of input current. It may appear that this problem is already solved in Figure 3A, which postulates neuronal response to the input current as one of the assumptions of our model. Indeed, in regions I and III (unconditionally OFF/ON) nothing can change the response postulated in Figure 3A. However, in the bistable region II the firing rate can follow one of the dependences, either ON or OFF. The problem is therefore to determine what branch of the firing rate dependence is followed after ON and OFF saccades.

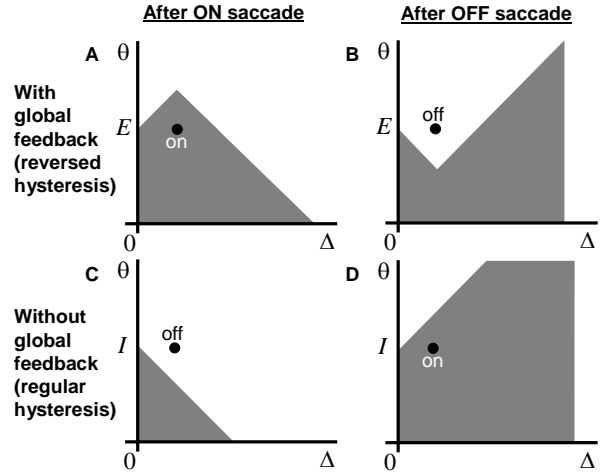

**Figure 8.** Reversal of hysteresis in the model with differential parameters. The response of a neuron (black dot) after the ON saccade (A) is higher than after the OFF saccade (B) when the recurrent connections are present. This is because this neuron is ON/OFF in these states as indicated. This is in contrast to the response with no hysteresis (C and D).

The qualitative difference between the case of independent neurons (Figure 3A, Results 1.1) and neurons connected by global feedback considered here is illustrated in Figure 8. Figures 8C and D show the activation function in the absence of feedback. Consider the neurons indicated by the black dot. For the same eye position after the ON/OFF saccades (Figure 8C) the neuron is OFF/ON, in agreement with Figure 3A. This is because the neurons are not connected and act independently. Such neurons therefore display regular direction of hysteresis. Figure 8A displays the activation function with the global feedback present. It is clear that for the same value of the input current and, consequently, the same eye position [see (18)] this neuron will be in the active state after the ON saccade. Similarly, after the OFF saccade, the neuron is expected to be OFF (Figure 8B). This behavior is different from the case of unconnected neurons (Figure 8C and D, 3A). This behavior is due to the drop of the recurrent current after the eye movement in the ON direction (increase after the OFF saccade) as was discussed in the previous paragraph. The presence of the global positive feedback therefore leads to the reversal of the sign of hysteresis for some neurons in the network.

We next examined, quantitatively, the dependence of the firing rate as a function of eye position in our model. We used the following method. Suppose that one has to calculate the firing rate after an ON saccade $f_{ON}(E)$. We evaluate the state of the neuron with the total input $E + \Delta I$ in the ON state (red dependence in Figure 3A) and subsequently follow the OFF firing rate dependence (blue in Figure 3A) from input $E + \Delta I$ to $E$. This latter operation reproduces the drop in the total recurrent current after the end of a saccade. A similar procedure is followed for the OFF saccades. The results of systematically applying this method are presented in Figure 9.

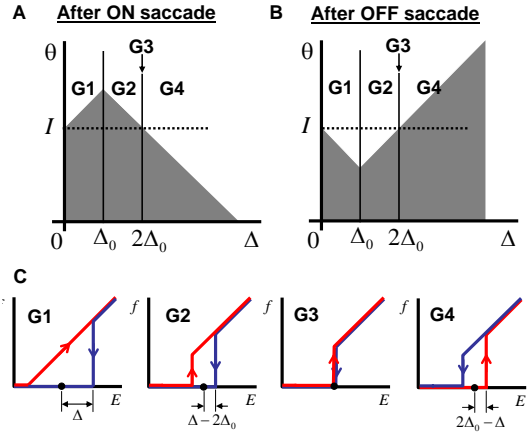

**Figure 9.** The firing rate as a function of eye position during fixations for different groups of neurons (G1-4). (A) and (B) the activation functions after the ON and OFF saccades respectively. The state of one neuron (black dot) is ON and OFF respectively. (C) Firing rate as a function of eye position for neurons from four groups (G1 to G4) indicated in (A). The red and blue curves correspond to the preceding ON and OFF saccades respectively.

Neurons can be separated into four groups with qualitatively different behaviors of the firing rates as functions of eye position. Figures 9A and B show the regions in the parameter space occupied by these groups with the corresponding dependencies of the firing rates on the eye position indicated in Figure 9C. For neurons with a large value of endogenous hysteresis (group G4) a regular direction of hysteresis is observed, the same as the endogenous one. This is not surprising, because the endogenous hysteresis is so strong for these neurons that no drop in the recurrent current can reverse it. The neurons with smaller values of endogenous hysteresis (parameter $\Delta$), which belong to groups G1 and G2 in Figure 9A, display a reversed hysteresis. Finally, a small group of neurons (G3) shows no hysteresis at all. We call these neurons marginal. The marginal neurons are important in establishing a relationship with our previous investigations[16] as explained below.

It is of interest to calculate the relative numbers of neurons belonging to different groups. This can be accomplished by using the conditions for areas occupied by different groups of neurons summarized in Figure 9A and equation (2). Thus, G1 contains 1/2 of all neurons, while G2 and G4 contain a 1/4 fraction each. Therefore, 3/4 of the neurons (G1 and G2) should demonstrate a reversed hysteresis, while for the 1/4 of the neurons (G4) the sign of the hysteresis should remain unchanged. These fractions are specific to the exponential distribution of the endogenous hysteresis $\rho(\Delta)$ given by (2). One can show that for some distributions $\rho(\Delta)$ the fraction of neurons with endogenous direction of hysteresis (G4) can be made arbitrarily small with respect to the cells with the reversed hysteresis direction (G1 and G2). Although the relative fractions of different groups of neurons (G1 versus G2) depend on the distribution of hysteresis widths (2), the firing rate dependences shown in Figure 9C do not. For a different distribution the neural responses as a function of eye position are exactly the same in this model.

The group G3 contains neurons with no hysteresis. This is despite the fact that without global feedback they should display a bistable response as in Figure 3A. This group is defined by the condition that their hysteresis half-width $\Delta$ is equal to $2\Delta_0$. Since this group resides on the interface between G2 and G4, the number of neurons in this group is small for the exponential distribution of $\Delta$. This however is not true for the case when all neurons have the same value of $\Delta$. We considered this case before in the present study [see (15)]. In this case *all* the neurons belong to the marginal group. This is because the position of these neurons in the parameter space $\Delta = 2\Delta_0$ determined by (16) coincides with the definition of the marginal neurons G3 in Figure 9A. Therefore in this case one should expect *no* hysteresis in the firing rate as a function of eye position displayed by all neurons. This conclusion applies to neurons in our previous study[16].

### 1.5 A more realistic implementation

We then examined if our simplified theory can apply to a more realistic model for a neural integrator involving biophysically plausible implementation of neuronal spiking and synaptic channels. We implemented our model with 40 two-compartmental neurons. Each neuron is represented by two compartments: somatic and dendritic (Figure 10A). The somatic compartment contains sodium and potassium conductances for the generation of action potentials. The firing frequency depends on the overall input into the dendritic compartment. The dendritic compartment is capable of generating membrane voltage-based bistability due to non-linearity pertinent to NMDA conductances. Since the amount of NMDA conductance is different for different cells (see Methods) the cells display a hysteretic input-output relationship in this system, similar to the one considered in the simple model. In particular, the firing rate of each cell as a function of input current into its dendritic compartment displays hysteresis.

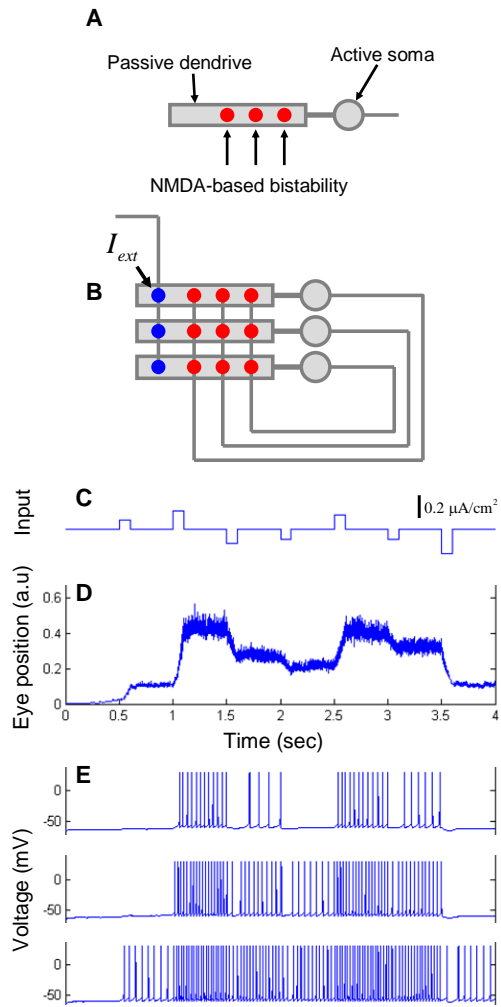

**Figure 10.** The model with Hodgkin-Huxley neurons. (A) Each neuron has a dendritic compartment that is bistable and soma that generates action potentials. To generate bistability the dendrite receives the tonic NMDA current. (B) Neurons are connected into the network by the all-to-all global feedback that generates additional NMDA current into each dendrite. The network contains 40 neurons, only three of which are shown. (C) An example of external input supplied to every dendritic compartment. (D) Resulting changes in the average feedback current reflect the integral of input. This variable could therefore be associated with the eye position. (E) The membrane voltage for three example neurons.

We then connected the cells by the recurrent feedback. The strength of all recurrent synapses is the same, thus implementing an all-to-all connectivity with roughly the same recurrent current on the input to all neurons (Figure 10B). When these cells are connected by recurrent feedback, the network can integrate a transient input current (Figure 10C-E) as a function of time.

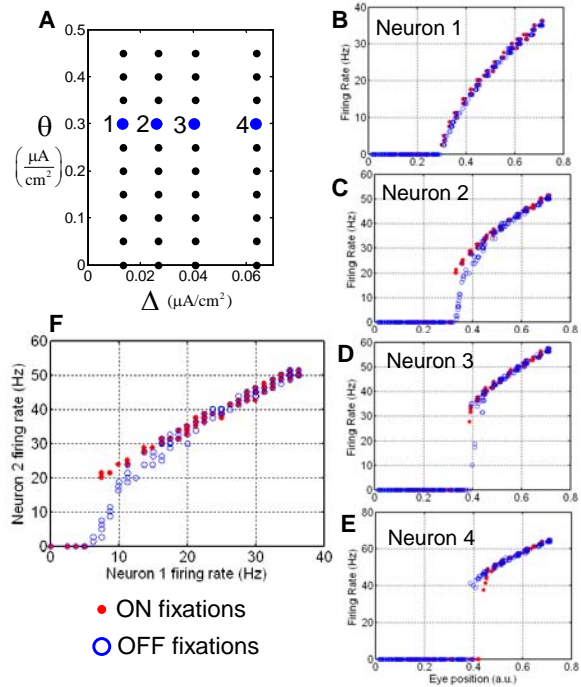

**Figure 11.** Reversal of hysteresis in the Hodgkin-Huxley neuron. (A) The arrangement of 40 neurons in the parameter space of hysteresis position and half-width. (B-E) The firing rate traces for four example neurons indicated in (A). The red/blue markers display values obtained after ON/OFF saccades. The red dependence is sometimes above the blue one (C and D) in agreement with the simple theory. (F) Firing rate of one neuron versus the other [(B) versus (A)] displays hysteresis, as observed experimentally.[17]

As predicted by the simple model, the neurons fall into four categories: with little or no hysteresis (Figure 11B), inverted hysteresis (Figure 11C), marginal neurons (Figure 11D), and regular direction of hysteresis (Figure 11E). These classes correspond to the dependences derived in the simplified model and illustrated in Figure 9C. We have therefore shown that the conclusions of the simplified model sustain the test by a more realistic implementation.

## 2. Multistable neurons.

We next examined the behavior of the recurrent system of neurons that themselves exhibit multistability. The cellular multistability is assumed to emerge from the bistable properties of several dendritic compartments (Figure 12A). We demonstrate here that under certain conditions the recurrent network of multistable neurons may be mapped mathematically onto the system of bistable neurons considered above. The properties of the multistable neurons in the network may be similar to the behaviors of bistable neurons, including the history-dependent profiles of the activation function (Figure 9) and the reversal of the sign of hysteresis.

We first consider the properties of multistable neurons in isolation, i.e. without recurrent connections present. Each neuron includes several dendritic compartments (Figure 12A) responsible for the generation of multistability. The dendritic compartments are assumed to be electrotonically isolated from each other. This implies that the response of each dendritic compartment is independent of the state of other compartments in the same dendritic tree and on the firing activity of the cell's soma. Each of the compartments is assumed to be bistable. The bistable ranges of the compartments are assumed to be distributed over a large range of values with no substantial difference in the hysteresis width. These assumptions lead to the dependence of the firing rate on the input current in the form of a staircase[15, 16]. This dependence is illustrated in Figure 12 B.

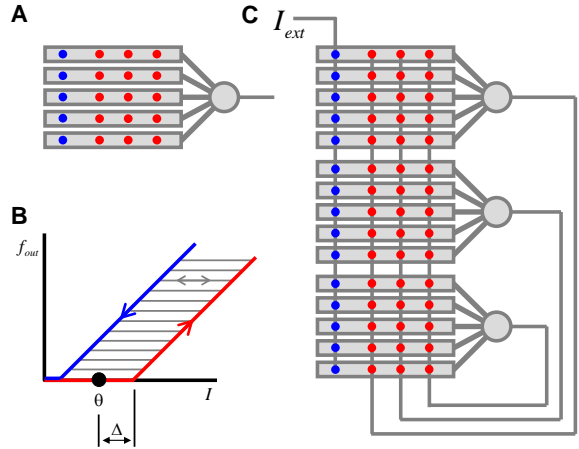

**Figure 12.** Model with multistable neurons. (A) Neuron with multiple bistable dendritic compartments. Synapses are shown by colored circles. (B) Firing rate as a function of input current for such a neuron is multistable: each stable firing state corresponds to the fixed number of active dendritic compartments. Introducing the parameters $\theta$ and $\Delta$ allows us to map the problem mathematically to the previous model with bistable neurons. (C) The recurrent network of multistable neurons with all-to-all connections that can be solved using this method.

Briefly, when the cell receives input current that increases, the dendritic compartments are sequentially switched on and the firing rate of the cell increases following the red line in Figure 12B. When the current decreases, the dendritic compartments are switched off leading to a decrease in the firing rate (blue line in Figure 12B). The space between the blue and the red lines corresponds to the bistable regime of the dendritic compartments. This implies that none of them will change their state and, therefore, the firing rate is preserved. This is signified by the gray horizontal segments in Figure 12B. The bistability in the behavior of dendritic compartments results in the hysteretic loop in the response of the cell[16, 19]. Note that the activation line (red) is below the deactivation line (blue) which means that for an isolated cell the hysteresis has a regular sign (cf. Figure 3). An additional twist is provided by the states that are accessible inside the hysteretic loop. We argue that the multistable neurons with these response properties connected into a recurrent network (Figure 12C) display reversed hysteresis through the mechanism similar to the previous discussion.

We now consider the multistable neurons connected into an all-to-all network (Figure 12C). As in our previous model we assume that all neurons receive the same value of input current composed of the external and recurrent currents. The latter component is determined by the number of neurons that are presently active. Similar to the previous system we assume that the neuron is active if the firing rate of this neuron is above zero. This assumption leads to the similarity between the model with multistable and bistable neurons (Section 1). Indeed, since the exact form of the firing rate dependence is not important for the network current, the neurons shown in Figure 3A and 12B are not distinguishable from the viewpoint of other neurons in the network. In particular, we can define each multistable neuron by two parameters: the median hysteresis position $\theta$ and half-width $\Delta$ (Figure 12B). The neurons then can be arranged on the 2D parameter plane similarly to the bistable case. Our conclusions about the dynamics of activation function can be transferred from the bistable case to the multistable case without any modification. Therefore, as far as the properties of the entire network are concerned, two networks based on bistable and multistable neurons are not distinguishable. The differences between the networks emerge when the firing rate of individual neurons is determined from the history dependence of the recurrent current. We illustrate this point next.

Consider a multistable neuron in the network. During the saccade in the ON direction the firing rate of this neuron was increasing according to the red dependence (Figure 13B). When the eye movement comes to conclusion the external input driving it terminates. The moment of termination of the external input is shown in Figures 13 A and B by point 1. Establishment of the eye position during fixations leads to a recoil in the recurrent current in this model by the amount equal to $2\Delta_0$, similar to the case of bistable neurons. From the point of view of the single neuron in the network, the input current is decreased by $2\Delta_0$ leading to the transition from point 1 illustrated in Figure 13B by the dotted arrow. The newly established firing rate during eye fixation is shown by the tip of the arrow. After the OFF saccade, the removal of external driving input leads to a symmetric increase in the recurrent current given by $2\Delta_0$ (Figure 13C). The corresponding firing rate for this neuron after the OFF saccade is represented by the tip of the dotted arrow in Figure 13D. Notice that for the same eye position in Figures 13B and D the firing rate is *higher* after an ON saccade than after the OFF saccade. This is in contrast to the response of the free neuron that is not connected by the feedback (Figure 12B). Thus, the hysteresis is expected to be reversed for this neuron due to the presence of network feedback.

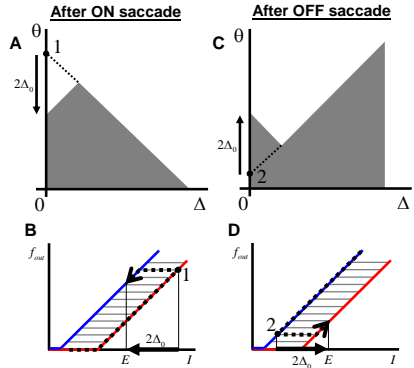

**Figure 13.** Illustration of the calculation of the firing rate as a function of eye position during fixations. For the same eye position *E* after ON saccade (A and B) the firing rate is larger than after the OFF saccade (C and D). This is similar to experimental findings (Figure 1).

We systematically applied the geometric procedure illustrated in Figure 13B and D to all groups of neurons in the parameter plane of the system. We obtained the dependence of the firing rate on the eye position illustrated in Figure 14. This shows that the neurons with substantially small hysteresis i.e. from groups G1 and G2, display inverted hysteresis as suggested by the qualitative analysis in Figure 13. For an inverted hysteresis the native hysteresis width $\Delta$ has to be smaller than the recoil in the recurrent current $2\Delta_0$. For neurons with strong native hysteresis from group G4, a regular sign of hysteresis is expected. These conclusions are similar to the results of the model with bistable neurons (Figure 9).

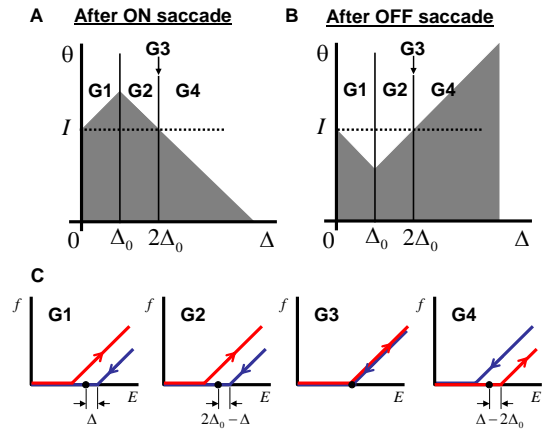

**Figure 14.** The firing rate as a function of eye position during fixations. The separation of the parameter space into areas (A and B) is similar to the bistable case (Figure 9). (C) The response is given by the inverted hysteretic loop.

The important feature displayed in Figure 14C is that the values of response inside the hysteretic loop are not accessible during fixations. This observation is an artifact of considering the saccades of large amplitude, which are independent of each other. If the actual eye movement included many smaller saccades, the values inside the hysteretic loop in Figure 14 would be possible. Thus one could make an experimental prediction that in the experimental data in Figure 1, smaller amplitude saccades result in the responses near the center of the hysteretic loop. Note also that the relative fraction of neurons in group G4, which are not observed experimentally, can be infinitesimally small, thus leading to a vanishing probability of encountering these neurons with regular hysteresis.

## 3. Mistuned integrator

In the previous sections we discussed the properties of the perfectly tuned integrator. In this case the current needed to recruit an additional neuron was exactly equal to the increase in the recurrent current resulting from recruiting this neuron. This condition is necessary for many states of the system to be equally stable, i.e. for the system to be multistable. The condition was quantitatively described by setting the parameter $\alpha = I_0 C \bar{\Delta}$ to 1. Here $I_0$ describes the contribution to the recurrent current from a single neuron and $C\bar{\Delta}$ is the inverse spacing between neuronal thresholds for activation, representing the increase in the recurrent current needed to activate an extra neuron. What if the recurrent feedback strength is weaker ($\alpha < 1$) or stronger ($\alpha > 1$)? We find that for substantial deviations from the perfectly tuned condition ($\alpha = 1$) stable eye fixations are also possible. Our model therefore displays the same degree of robustness that is pertinent to simpler hysteretic systems[15, 16]. The activation function during fixations is displayed for the case of weak feedback in Figure 15. Although the fixation is stable, the recoil in the recurrent current after the ON saccade is larger than the increase after the OFF saccade (Figure 15A versus B). The activation function therefore loses its symmetry between ON and OFF saccade cases pertinent to the perfectly tuned integrators (Figures 9 and 13).

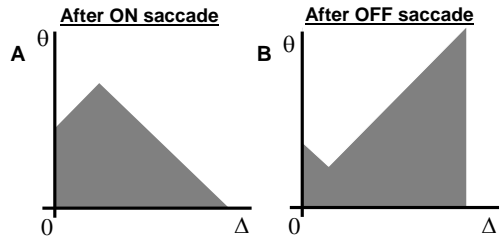

**Figure 15.** Robustness of a mistuned integrator. Mistuning $\alpha = I_0 C \bar{\Delta} \neq 1$ [defined in (10)] leads to the asymmetric activation function after the ON and OFF saccades. The activation functions shown correspond to $\alpha < 1$, i.e. the weak feedback.

## 4. Subthreshold inputs

Our model includes some neurons that have no hysteresis ($\Delta = 0$). At the same time we have reported above that sustained integration requires the external input exceeding a certain threshold equal to the average value of hysteresis for the ensemble (Figure 5). The natural question is whether the neurons with no hysteresis can somehow integrate even a subthreshold input ($I_{ext} < \bar{\Delta}$). 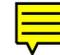

In the model described above the subthreshold inputs can indeed induce persistent changes in the activation function and the firing rates of integrator neurons. It is true, however, that these changes will not depend on the duration of the stimulus. They are only determined on the stimulus amplitude. Thus, persistent integration is indeed not possible, which means that the longer stimulus will not induce larger changes in the number of active neurons. This statement is valid for the model with no noise, i.e. with no spontaneous switching of the bistable units. We demonstrate in below that the presence of a finite switching time will lead to the leaky integrator and will make sustained integration possible even for a subthreshold inputs.

The positive input, ($I_{ext} > 0$), leads to no sustained changes in the activation function. This implies that when the input is extinguished (Figure 16C) the activation function is the same as before the stimulus (Figure 16A). This is because the stability equation (13) has only one solution in this case as shown in Figures 16A and C. However, the negative inputs, ($I_{ext} > 0$), do lead to a sustained decrease in the eye position after an ON saccade (Figure 16D-F). This partially justifies the intuition that the units with small hysteresis can react to stimuli of small amplitude. Of course this intuition fails for positive inputs (Figure 16A-C), since they do not lead to sustained changes in the integrator state.

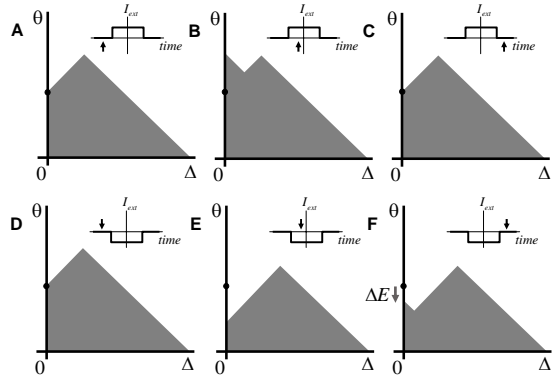

**Figure 16.** Response of the integrator to the subthreshold inputs $I_{ext} < \overline{\Delta}$ is asymmetric. After the ON saccade, the inputs in the same direction lead to *no* sustained changes in the activation function (A-C). The inputs in the opposite direction results in a sustained decrease in the eye position (D-F) shown by the gray arrow. The initial eye position and the moment of time are indicated by the full circle and the arrow in the inset.

In the velocity-to-position integrator the asymmetry in the response to the subthreshold inputs can be compensated by the use of two coupled integrators, on both sides of the brain. Their complimentary sensitivity to the subthreshold inputs may resolve the potential problem with integration of the small inputs. Another possibility is the integrator leak that is described in the next section.

## 5. Integration time-constants

In this section we study the effect of spontaneous transitions in the bistable neurons on the stability of the integrator as a whole. So far we have assumed that the bistable or multistable neurons possess the property of perfect memory, i.e. there are no transitions between different states for these neurons. Thus, in the case of bistable neurons we assumed that the neuron will remain in the ON or OFF state for a very long time, even though a transition to the other state is possible. In the case of multistable neurons we assumed the state of bistable dendritic compartments is preserved, which leads to the infinite in time retention of the value of the firing rate for fixed inputs. In this section we examine the effects of noise-driven transitions between different memory states. We will address the behavior of bistable neurons for definiteness. We derive the connection between the decay time-constant of the bistable units and the integrator leak time.

In case of perfect memory the activation function $h(\Delta, \theta)$ can take two values, 1 and 0, corresponding to the ON and OFF states respectively. When the spontaneous noise-driven transitions are possible, the activation function may take values between 0 and 1. This is because the activation function defines the average state of neurons in the small neighborhood of a point in the parameter space. If spontaneous transitions are allowed, the average value can deviate from pure values of 0 and 1. The activation function $h(\Delta, \theta)$ then describes the fraction of neurons in the ON state near point $(\Delta, \theta)$. It is also equal to the probability of finding a unit in the active state. The evolution of this function can be described by the relaxation equation

$$\frac{\partial h(\Delta, \theta)}{\partial t} = \frac{h_0(\Delta, \theta) - h(\Delta, \theta)}{\tau(\Delta, \theta)}. \qquad (19)$$

Here $h_0(\Delta, \theta)$ is the activation function in the equilibrium, while $\tau(\Delta, \theta)$ is the relaxation time-constant describing how fast this equilibrium is reached. The important feature of this equation is that both the equilibrium relaxation function and the time constant depend on the position in the parameter space. Indeed, in an area of unconditional stability [I and III in Figure 3C], the equilibrium function equals 0 and 1 respectively. No other values are permitted because there is no bistability in these regions. In the area of bistability [II in Figure 3C], the equilibrium function takes intermediate values that are determined by the relative fraction of units in the ON and OFF states. Also, in areas I and III the equilibrium values are reached with a very fast time constant. We assume that the bistable units can be flipped essentially instantaneously. In the area of bistability [area II], the equilibrium is reached over much longer time-scales determined by the time of spontaneous transitions between two stable states of the neuron. Thus if one assumes that the transition from the ON state to OFF state (decay) is characterized by the time constant $\tau_{1\to 0}$ while the opposite transition (spontaneous activation) occurs with time-constant $\tau_{0\to 1}$, the equilibrium activation function and the time-constant of relaxation to the equilibrium value in area II is

$$h_0 = \tau_{1\to 0}/(\tau_{1\to 0} + \tau_{0\to 1}) \qquad (20)$$

$$\tau_h = \tau_{0\to 1}\tau_{1\to 0}/(\tau_{1\to 0} + \tau_{0\to 1}). \qquad (21)$$

To simplify the notations we assume that $\tau_{0\to 1} = \tau_{1\to 0}$ in the subsequent consideration, so that in area II $h_0 = 1/2$ and $\tau_h = \tau_{1\to 0}/2$. The important feature of the biological system is that the decay time-scale $\tau_h$ is substantially larger than synaptic time-constant $\tau_s$. The ratio between $\tau_h$ and $\tau_s$ is usually exponential[20, 21]. The assumption that bistability can be maintained longer than synaptic time scale implies that the recurrent current can circulate in the system several times before the bistability decays. This assumption allows us to substantially simplify solving (19).

Just to demonstrate that the seemingly simple (19) can lead to interesting results, we solved the equation for the case when $\tau(\Delta,\theta) = \tau_h$ is constant everywhere in area II. Our goal is to understand how the decay of bistable neuronal units translates into the decay of integrator memory, or, in other words, how the integrator becomes leaky. That the integrator is leaky implies that the recurrent current changes with time. For example, if the recurrent connections are weak ($\alpha < 1$) the current decays ($dI/dt \equiv \dot{I}$ is negative). The equilibrium activation function in this case is sliding down in the parameter space at a rate equal to $\dot{I}$ as illustrated in Figure 17A. Note that the equilibrium activation function is the limit towards which the real activation function is moving at each moment in time, according to (19). The equilibrium activation function therefore can be non-stationary, in which case the final target for the activation function is dynamically changing. Thus, for $\dot{I} < 0$ (Figure 17), the real activation function, lags behind the equilibrium values by the time constant $\tau_h$. Solution for the activation function in area II becomes

$$h = 1/2 + e^{-\Delta t/\tau_h}/2. \qquad (22)$$

Here

$$\Delta t = (\theta - I + \Delta)/|\dot{I}| \qquad (23)$$

is the duration of time spent by the neuron at point $(\Delta,\theta)$ in area II. The number of active neurons can be easily calculated from (22) using (6). We obtain

$$n = C\overline{\Delta}I + \frac{C}{2}\int_0^\infty d\Delta \int_{I-\Delta}^{I+\Delta} d\theta e^{-(\Delta+\theta-I)/\tau_h|\dot{I}|-\Delta/\overline{\Delta}} =$$

$$= C\overline{\Delta}I - C\overline{\Delta}\tau_h \dot{I}\frac{\overline{\Delta}}{2\overline{\Delta} + \tau_h|\dot{I}|} \qquad (24)$$

This equation is valid for an arbitrary sign of the time-derivative of the current $\dot{I}$: both positive and negative. Note that for $\dot{I} < 0$ the correction to the number of active neurons [second term in (24)] is positive. This is because the activation function 'extends' into area II due to a finite delay time introduced by hysteresis (gray plume in area II in Figure 17B).

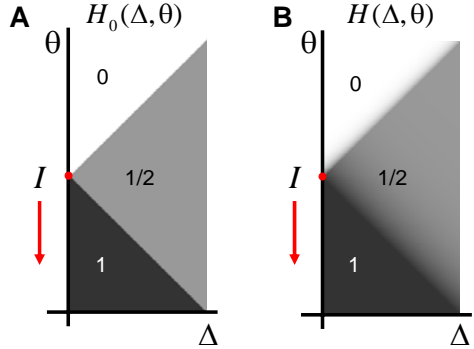

**Figure 17.** The activation function in the case of non-stationary recurrent current that decreases as a function of time. (A) The equilibrium activation function for the case $\tau_{0\to1} = \tau_{1\to0}$ (B) The actual activation function is lagging behind the equilibrium values.

To simplify analysis further we assume that the rate of change of the current is actually small. This is possible if the integrator is tuned. The more exact condition for this assumption to be valid becomes clear in the following equations. Assume that $\tau_h |\dot{I}|$ is much smaller than $\overline{\Delta}$ in the denominator in (24). Under this condition we can write that

$$n(t) = C\overline{\Delta} \cdot I(t) - C\overline{\Delta} \cdot \tau_h \dot{I}/2. \qquad (25)$$

Since the recurrent current is proportional to the number of active neurons in this model

$$I(t + \tau_s) = I_0 n(t) \qquad (26)$$

we then arrive at the equation describing the dynamics of input current to all units:

$$(1-\alpha)I = \frac{\alpha \tau_h}{2} \frac{dI}{dt}. \qquad (27)$$

Here the tuning parameter $\alpha = I_0 C\overline{\Delta}$ is the same as defined above. It is clear from this equation that the mistuning of the integrator should be small, i.e. $(1-\alpha)I \ll \overline{\Delta}$, for us to neglect the second term in the denominator of (24). Thus the leaky integrator equation is valid if the mistuning is not too large.

Equation (27) describes decay of the integrator current in the absence of the external inputs. The integrator time-constant that follows from this equation is

$$\tau_{\text{leak}} = \frac{\tau_h}{2|1-\alpha|} \qquad (28)$$

We conclude that the integrator leak time-constant is determined by the rate of decay of bistability in this model. This is in contrast to the models without substantial hysteresis in which the time-scale for the integrator leak is provided by the synaptic time-constant $\tau_s$ [10, 13]. This point suggests another interpretation for robustness of this model. Indeed, since observed $\tau_{\text{leak}}$ is about 30 sec, with the time-scale of the decay of hysteresis $\tau_h$ of a few seconds[20, 21], the parameter $\alpha$ has to be tuned to unity with the precision of about 10%. Thus the presence of hysteretic neurons allows putting a much weaker constraint on the integrator tuning to reach the same value of leak.

With the help of (24) one can derive a more general equation for the dynamics of the integrator:

$$\left(1-\alpha\right)I+\left(\frac{\alpha\bar{\Delta}\tau_h}{2\bar{\Delta}+\tau_h\,|\,dI/dt\,|}+\tau_s\right)\frac{dI}{dt}=I_{ext} \quad (29)$$

This equation is valid for a constant external current. For a perfect tuning condition ($\alpha=1$) and large external current $I_{ext}$ this equation becomes

$$\tau_s\frac{dI}{dt}=I_{ext}(t)-\bar{\Delta}\cdot\text{sign}\left(\frac{dI}{dt}\right). \quad (30)$$

This equation yields solution (11) previously obtained. Another point evident from this equation is that the rate of integration is determined by the synaptic time-scale $\tau_s$, while the decay is proportional to the average time of spontaneous decay of the bistability $\tau_h$. The latter is an exponential function of the former[20, 21], and may significantly exceed the synaptic time-scale. Thus, the neural integrator described here is capable of integrating the external stimuli effectively due to a small time-constant $\tau_s$ while remaining robust due to the larger memory time-scale $\tau_h$.

Finally, we note that for the perfectly tuned condition ($\alpha=1$) the current in the integrator is changing ($dI/dt\neq 0$) even for infinitely small external current. This implies that the integrator with spontaneous transitions does not have the threshold for integration. The rate of integration of weak inputs is determined however by the hysteresis time-constant $\tau_h$, which is a remnant of the threshold existing for perfect bistable units without spontaneous transitions between the ON and OFF states (Figure 5).

## 6. Experimental predictions

The main experimental prediction that follows from this model concerns the comparison between the firing rates during fixations and slow eye movements. The latter include both smooth pursuit and vestibule-ocular reflex (VOR) responses. The previous sections have demonstrated that the model presented predicts an inverted sign of hysteresis for some neurons in the ensemble, similarly to what is observed experientially. Thus ON responses are above OFF responses in Figures 9 and 14 for G1 and G2. We argue here that the responses of the same neurons are different during smooth eye movements, sometimes in a qualitative manner.

To present our argument we notice that the activation function is the same for the smooth eye movements as for the case when the neurons are not connected by global recurrent connections (compare Figures 4A and 5A). Thus, the firing rate as a function of eye position can be derived directly from the dependence with no recurrent connections by choosing the appropriate value of input current. Thus, the total value of the input current is composed of the recurrent and external current. We have associated the former with the eye position (18). The external current has a minimum value that allows sustained integration, at least, if decays in bistability are ignored. We thus obtain for the smooth eye movements

$$I=E+\bar{\Delta}\cdot\text{sign}(I_{ext}) \quad (31)$$

The firing rate as a function of eye position is obtained from the dependences with no feedback by shifting the ON dependence to the left by $\bar{\Delta}$ and the OFF dependence to the right by $\bar{\Delta}$. The dependence is shown in Figure 18B for the multistable neurons. As shown in Figure 18, the neurons in group G1 with small hysteresis will acquire history dependence during eye movements. This feature could be detected experimentally.

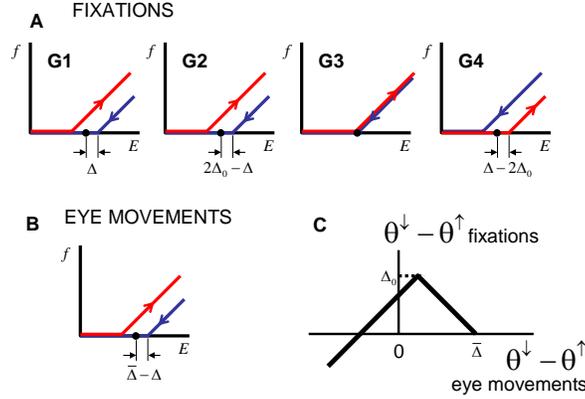

**Figure 18.** The differences in responses between the case of eye fixations (A), and smooth movements (B), predicted by this model. (C) The distance between OFF and ON thresholds for fixations versus small velocity eye movements that follows from (A) and (B).

## Discussion

We studied the recurrent networks built out of hysteretic units. In contrast to previous studies we considered the case of neurons having different values of hysteresis. We analyzed the properties of such networks assuming a large number of neurons, which allowed treating the behavior of the system statistically, using distribution functions. The particular assumption made here is that the neurons are connected in the all-to-all fashion, i.e. each neuron makes synapses with the dendrites of all other neurons in the network. This assumption allowed us to study the properties of this network analytically, i.e. without the use of a computer, and to prove many statements exactly.

One of the important properties derived here is that the hysteresis of the neurons is strongly affected by the recurrent connections. Assume that before neurons are connected, their firing rate exhibited history dependence of a particular sign. For example, assume that the firing rate would always be higher after the decrease in the external input into the neuron than after an increase. Such history dependence is often displayed by neurons with intracellular mechanisms of positive feedback, or in local strongly-connected clusters of neurons[16]. We called, somewhat arbitrarily, this type of history dependence a regular native hysteresis or a hysteresis of positive sign. As follows from our consideration, the history dependence after these neurons are connected in the network is completely different. Many neurons actually change the sign or direction of hysteresis from positive (regular) to negative (anomalous). This implies that the firing rates after increase in the external current (ON saccades) is higher than after a decrease in the current. This feature is acquired by the neurons when they are connected into the network, which is needed to maintain the memory about the eye position. Similarly, it is quite easy to show that the neurons with negative native hysteresis may reverse their sign to positive when connected into networks.

This phenomenon of the reversal of the direction of history dependence in the recurrent networks of hysteretic neurons could reconcile the observed direction with the possible mechanisms of generating hysteresis in the VPNI. Indeed, the observed sign of hysteresis is almost always negative[17]. Simplistically, this may imply that the mechanisms other than those involving positive feedback are responsible for this sign of hysteresis, such as the mechanisms based on negative feedback. Our finding here indicates that the neurons could have a positive feedback active inside the cells generating the regular direction of history dependence, which, after the neurons are connected into the network, becomes reversed. In a sense, two positive feedbacks, intracellular and extracellular, may manifest themselves in reversed hysteresis mimicking the negative feedback systems.

We also studied the effect of spontaneous transitions of the bistable units on the dynamics of the integrator. We found that in the presence of such transitions the integrator becomes leaky. The leak time-constant is determined by the rate of spontaneous transitions. We derived the simple formula relating two time constants, those of integrator leak and of spontaneous transitions [see (28)]. The relationship is similar to the earlier derived formula for the connection between the fast synaptic time-constant and integrator leak for integrators with no hysteresis. The additional twist is that the actual integration rate for the hysteretic integrator is determined by the faster synaptic time-constant. Thus, hysteretic integrators are both robust and capable of responding quickly to the external stimuli.

Our consideration of the dynamics of transitions in the hysteretic neurons was limited for two reasons. First, we assumed that the transition time-constant is the same for all units, independent of the width of hysteresis. Second, we assumed that the spontaneous transition rate is the same at all positions in the region of bistability. Our approach allows relaxing these assumptions at the expense of simplicity. We argue that making small hysteresis units less robust to noise-driven transitions, as predicted by prior theoretical studies[16, 20], will allow integrating small amplitude inputs. The systems with variable spontaneous transition rates therefore deserve further investigation.

Our consideration of network dynamics was limited to the evolution of the distribution function in the two-dimensional parameter space (hysteresis width and position). Additional parameters could be included in consideration by making the coordinate space for the distribution function three, or more than three dimensional. Such variables would allow a study of network topologies more complex than all-to-all. Indeed, an additional parameter or parameters could specify the locus of each neuron in the network.

## Conclusion

We studied an exactly solvable model for recurrent networks of hysteretic neurons. This model displays the reversal of the direction of hysteresis when the recurrent connections are included. The leak time-constant of the integrator with hysteretic neurons is determined by the rate of spontaneous noise-driven transitions in the individual neurons. We argue that experimental data are consistent with the positive feedback existing on both intracellular and network levels.

## Appendix A

Here we apply the stability condition (13) to some distributions of hysteresis widths $\rho(\Delta)$ and obtain the parameter of stable configuration $\Delta_0$ (Figure 6B). The boundary function in Figure 6B is

$$b(\Delta) = \begin{cases} \Delta, & \Delta < \Delta_0 \\ 2\Delta - \Delta, & \Delta \geq \Delta_0 \end{cases} \quad (32)$$

The goal is to find $\Delta_0$ using equation (13). Using (32), equation (13) can be rewritten as follows

$$\int_0^{\Delta_0} \rho(\Delta)\Delta d\Delta + \int_{\Delta_0}^{\infty} \rho(\Delta)(2\Delta_0 - \Delta)d\Delta = 0 \quad (33)$$

For the exponential distribution $\rho(\Delta)$ given by equation (2) from (33) we obtain

$$\Delta_0 = \overline{\Delta} \ln 2 \quad (34)$$

For the delta-function distribution of hystereses (15), $\Delta_0 = \overline{\Delta}/2$, which can be verified by direct substitution to (33). This leads directly to (16).